\documentclass[aps,prr,twocolumn,superscriptaddress,showpacs,amsmath,amssymb, amsfonts,10pt]{revtex4-1}

\usepackage{graphicx, placeins}
\usepackage[unicode=true,pdfusetitle,bookmarks=true,bookmarksnumbered=true,bookmarksopen=false, breaklinks=false,pdfborder={0 0 0},pdfborderstyle={},backref=false,colorlinks=false, pdftitle={Probing molecular spectral functions and unconventional pairing
using Raman spectroscopy}]{hyperref}
\usepackage{amsmath}
\usepackage{lipsum}  
\usepackage{color}
\usepackage{verbatim}
\usepackage{txfonts}
\usepackage[T1]{fontenc}

\usepackage[normalem]{ulem} 

\definecolor{black}{rgb}{0,0,0}
\definecolor{blue}{rgb}{0,0,1}
\definecolor{green}{rgb}{0,1,0}
\definecolor{red}{rgb}{1,0,0}
\definecolor{brown}{rgb}{0.4,0.2,0}
\definecolor{darkgreen}{rgb}{0,0.7,0}
\definecolor{darkblue}{rgb}{0.0,0.0,0.5}
\definecolor{red}{rgb}{1,0,0}
\definecolor{deepmagenta}{rgb}{0.8, 0.0, 0.8}

\newcommand{\pv}{\mathbf{p}}
\newcommand{\vecp}{\mathbf{p}}

\newcommand{\vecq}{\mathbf{q}}
\newcommand{\kv}{\mathbf{k}}
\newcommand{\qv}{\mathbf{q}}
\newcommand{\lv}{\mathbf{l}}

\newcommand{\Io}{\hat{\mathbf{I}}}
\newcommand{\So}{\hat{\mathbf{S}}}
\usepackage{braket}
\usepackage{appendix}
\usepackage{ulem}

\begin{document}
\title{Probing molecular spectral functions and  unconventional  pairing \\ using Raman spectroscopy}

\author{Oriana K. Diessel}
\affiliation{Max-Planck-Institute of Quantum Optics, Hans-Kopfermann-Strasse 1 , 85748 Garching, Germany}
\affiliation{Munich Center for Quantum Science and Technology (MCQST), Schellingstr. 4, 80799 Munich, Germany}

\author{Jonas von Milczewski}
\affiliation{Max-Planck-Institute of Quantum Optics, Hans-Kopfermann-Strasse 1 , 85748 Garching, Germany}
\affiliation{Munich Center for Quantum Science and Technology (MCQST), Schellingstr. 4, 80799 Munich, Germany}
\affiliation{Department of Physics, Harvard University, Cambridge, Massachusetts 02138, USA}

\author{Arthur Christianen}
\affiliation{Max-Planck-Institute of Quantum Optics, Hans-Kopfermann-Strasse 1 , 85748 Garching, Germany}
\affiliation{Munich Center for Quantum Science and Technology (MCQST), Schellingstr. 4, 80799 Munich, Germany}

\author{Richard Schmidt}
\affiliation{Max-Planck-Institute of Quantum Optics, Hans-Kopfermann-Strasse 1 , 85748 Garching, Germany}
\affiliation{Munich Center for Quantum Science and Technology (MCQST), Schellingstr. 4, 80799 Munich, Germany}
\affiliation{Center for Complex Quantum Systems, Department of Physics and Astronomy, Aarhus University, 8000 Aarhus C, Denmark}
\affiliation{Institute for Theoretical Physics, Heidelberg University, Philosophenweg 16, 69120 Heidelberg, Germany}

\date{\today}
\begin{abstract}
An impurity interacting with an ultracold Fermi gas can form either a polaron state or a dressed molecular state in which the impurity forms a bound state with one gas particle. 
This molecular state features rich physics, including a first-order transition to the polaron state and a negative effective mass at small interactions. However, these features have remained so far experimentally inaccessible. In this work we show theoretically how the molecular state can be directly prepared  experimentally even in its excited state using state-of-the-art cold atom  Raman spectroscopy techniques. Initializing the system in the ultra-strong coupling limit, where the binding energy of the molaron is much larger than the Fermi energy, our protocol maps out the momentum-dependent spectral function of the molecule.
Using a diagrammatic approach we furthermore show that the molecular spectral function serves as a direct precursor of the elusive Fulde-Ferell-Larkin-Ovchinnikov phase, which is realized for a finite density of fermionic impurity particles. Our results pave the way to a systematic understanding of how composite particles form in quantum many-body environments and provide a basis to develop new schemes for the observation of exotic phases of quantum many-body systems. 
\end{abstract}

\maketitle
Understanding the nature  of composite particles in a quantum medium is essential to unveil the physics of many intriguing phases of matter. Notable examples include Cooper pairs in superconductors \cite{Cooper_1956,Bardeen_1957}, the BEC-BCS crossover in ultracold gases \cite{Eagles_1969,Leggett_1980,Nozieres_1985,Regal_2004,Zwierlein_2004,Chin_2004,Kinast_2004,Bourdel_2004,Strecker_2003},  superfluids of excitons in semiconductors \cite{Blatt_1962,Gergel_1968}
, anyons such as flux-tube-particle composites \cite{Wilczek_1982},  and the composite baryons and mesons arising from the quark-gluon plasma in the QCD phase diagram \cite{Cabibbo_1975}. 

A paradigmatic system to understanding the formation of such composite particles in a quantum environment is the Fermi polaron problem.
Here one distinguishable particle (a `quantum  impurity') interacts attractively with a bath of indistinguishable fermions. As the attraction between the impurity and the bath increases,  a first-order transition is predicted to occur  between a polaron state, in which the impurity is dressed by 
bath fluctuations, and a composite molecule state in which the impurity is  bound to  one of the bath fermions \cite{Chevy2006,Prokofev_2008,Prokofev2008a,Mora_2009,Punk_2009,Combescot_2009,Cui2010,Bruun_2010,Schmidt_2011,massignan2011repulsive,Trefzger_2012,Kohstall_2012}.
For strong attraction the molecule is a tightly bound composite that is nearly unaffected by the quantum medium. However,  as the transition is approached the molecule experiences dressing by and exchange with the bath fermions and forms a \textit{molaron}~\cite{SchirotzekThesis}, a composite quasiparticle.

\begin{figure}[t]
	\centering
	\includegraphics[width=
	\linewidth]{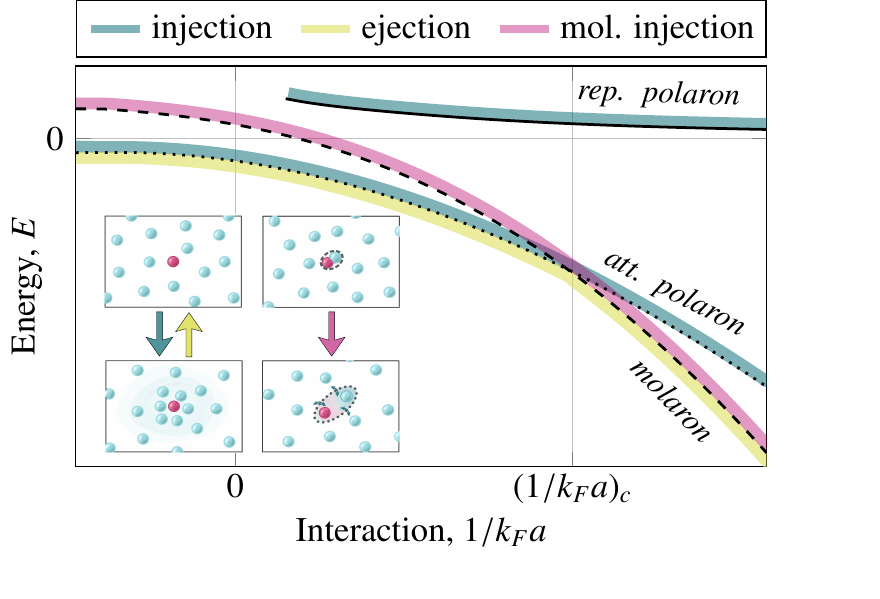}
	\caption{Schematic figure of the polaron (dotted black line) to molaron (dashed black line) transition. The inset displays the different spectroscopic schemes in ultracold quantum gases: In injection spectroscopy (blue arrow), an initially non-interacting impurity is injected into a state in which it can interact with the bath particles. This technique allows for the detection of the polaron, but suffers from a vanishing overlap between the non-interacting ground state and the molaron state. In ejection spectroscopy (yellow arrow), the interacting impurity gets ejected into a non-interacting state, which allows for the detection of the ground state (yellow). 
	Our proposal (pink arrow) enables the detection of the excited molaron branch, by initializing the system in the ultra-strong coupling limit.
}
	\label{fig:NewFig1}	
\end{figure}

Understanding the formation of molarons and their properties in many-body environments is essential to fully describe the general phase diagram of imbalanced Fermi mixtures in ultracold quantum gases and neutron matter~\cite{Alpar_1995}, as well as the physics of trions in doped, atomically thin semiconductors~\cite{Sidler_2017}. 
However, so far molarons have remained experimentally inaccessible.
One of the reasons is that typical probes in condensed matter physics act on the single particle level, relying for instance on tunneling of single electrons in solids~\cite{Binnig1982,Binnig1987} or electronic transitions of single atoms in cold atom experiments \cite{Schirotzek_2009,Kohstall_2012,Ness_2020,Fritsche_2021,Scazza_2022,Yan_2019,Vale_2021,Liu_2020,Liu_2020_2,Wang_2022,Wang_2022_2} (cf. Fig.~\ref{fig:NewFig1}). Such limitations hinder the direct creation of the composite particles and have so far precluded the spectroscopy of molarons including their momentum-resolved excitation spectrum.

In this letter we propose a new spectroscopic protocol based on Raman transitions to probe composite quasiparticles in cold atomic many-body systems. Using this new  technique allows one to directly measure the momentum-resolved molaron spectral function that encodes all information about the composite quasiparticle including its dispersion, lifetime,  effective mass, and full excitation spectrum. Furthermore, using a  diagrammatic resummation method, we demonstrate that the finite-momentum properties of the excited molaron state are intimately connected to the  emergence of the Fulde-Ferell-Larkin-Ovchinnikov (FFLO) \cite{Fulde_1964,Larkin_1964} phase at finite impurity density, where  composite Cooper pairs condense into a finite-momentum state. Our finding highlights a remarkable connection between  unconventional superconductivity  and  Fermi polarons that allows one to observe fingerprints of complex many-body phases in  quantum impurity problems. 
\begin{figure*}[t]
	\centering
	\includegraphics[width=
	\linewidth]{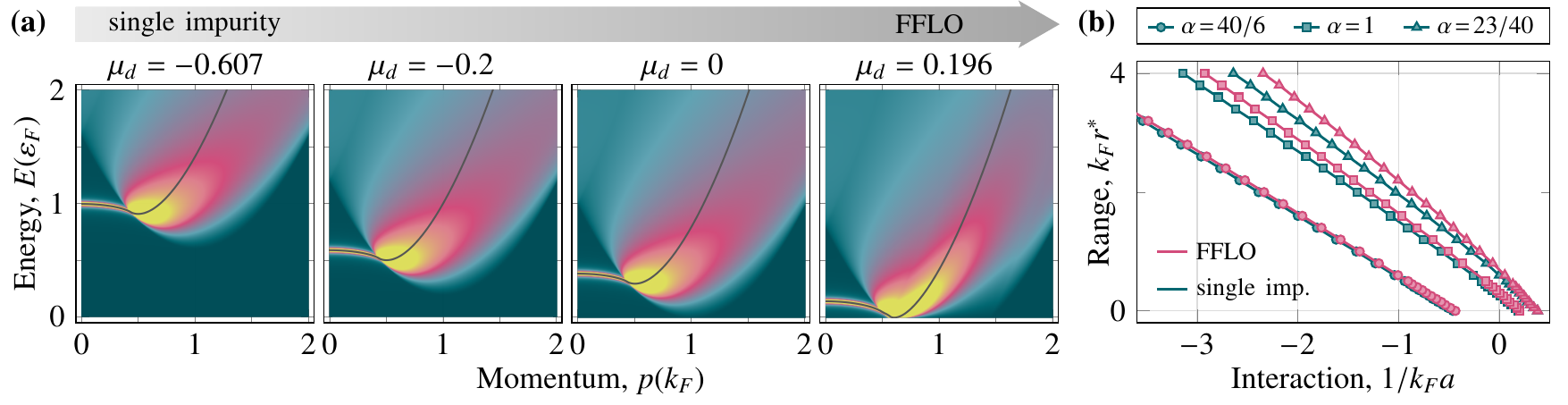}
	\caption{(a) Molaron spectral function at unitarity in the contact interaction limit ($k_Fr^*\!=\!0$) for different chemical potentials of the minority population. The molaron dispersion (gray line) has a minimum at finite momentum. For $\mu_d=0.196\, \epsilon_F$, the molaron becomes gapless at finite momentum, which determines the onset of FFLO. (b) Critical range parameter $k_Fr^*$  below which the molaron dispersion minimum is at finite momentum (blue) and for which FFLO sets on at finite impurity density (pink) for different mass ratios $\alpha$ as function of the interaction strength.}
	\label{fig:Fig1}	
\end{figure*}

\textit{Microscopic model.---} We start by introducing a two-channel Hamiltonian~\cite{Bruun_2004} to model the Fermi polaron problem close to a Feshbach resonance
\begin{align}
\mathcal{H}=&\sum_{\pv}\varepsilon_{\pv}^{c}c_{\pv}^{\dagger}c_{\pv}^{\phantom{\dagger}}
+
\sum_{\pv}\varepsilon_{\pv}^{d}d_{\pv}^{\dagger}d_{\pv}^{\phantom{\dagger}}
+
\sum_{\pv}\left(\xi_{\pv}+\nu\right)m_{\pv}^{\dagger}m_{\pv}^{\phantom{\dagger}}
\nonumber\\
&+\frac{h}{\sqrt{V}}\sum_{\lv,\pv}m_{\pv}^{\dagger}c_{\lv}^{\phantom{\dagger}}d_{-\lv+\pv}^{\phantom{\dagger}}+\text{h.c.} \ .
\label{Eq:2-Ch_Hamiltonian}
\end{align}
Here an impurity ($d^{\dagger}_\pv$) interacts with a bath of $N$ fermions ($c^{\dagger}_\pv$) in a volume $V$ via the exchange of a bare molecular state in a closed scattering channel ($m^{\dagger}_\pv$). The dispersion relations of the impurity and the bath particles are given by $\varepsilon^{d/c}_\pv=\pv^2/2M_{d/c}$, respectively, with momentum $\pv$ and masses $M_c$ and $M_d\equiv\alpha M_c$. The molecule has a dispersion relation $\xi_{\pv}= \pv^2 /2 (1+ \alpha) M_c$ and an energy detuning $\nu$ which in experiments can be tuned by a magnetic field.

The interaction in the system is described by the last term of Eq.~\eqref{Eq:2-Ch_Hamiltonian}, where two fermions in the open scattering channel are converted into the closed-channel molecule.
The conversion factor $h$ is proportional to the width of the Feshbach resonance and along with the detuning $\nu$ it is tuned to reproduce the  $s$-wave scattering length $a$ and range parameter $r^*$ of the impurity-bath interaction  via $\nu_0/h^2=-\mu/2 \pi a+V^{-1} \sum_\kv 1/(\varepsilon^c_{\kv}+\varepsilon^d_{\kv})$~\footnote{$r^*$ is related to the effective range $r_e$ via $r^*=-r_e/2$} and $h^2=\pi/r^* \mu^2$ with $\mu=M_dM_c/(M_d+M_c)$~\cite{chin:2010}.

At finite fermion density (determined by the Fermi wave vector $\kv_F$) the impurity is dressed by  fluctuations in the Fermi gas and forms a polaron. The molecule in the model is also dressed and forms a composite quasiparticle, the \textit{molaron}. This molaron is adiabatically connected to the molecular bound state at strong coupling.
Theoretically both the polaron and molaron can be described using diagrammatic techniques, where the central object is the  retarded Green's function
\begin{align}\label{GPolDef}
    \!\!G^R(E,\vecp) = \mathcal{F} \left[ -i\Theta(t)\left\langle\text{FS}\left|\left[X_{\pv}^{\textcolor{white}{l}}(t),X_{\pv}^{\dagger}(0)\right]_{\pm}\right|\text{FS}\right\rangle \right]\!
\end{align}
with $X_\pv=d_\pv (m_\pv)$ for the impurity (molecule) and $[.\, ,.]_{\pm}$ the (anti-)commutator.
Here $\mathcal{F}$ denotes the Fourier transform from time $t$ to frequency space $E$. The poles of $G^R(E,\vecp)$ directly yield the  energy of the attractive and repulsive polaron (molaron) as schematically shown in Fig.~\ref{fig:NewFig1}. 
As can be seen in the figure, the molaron also exists as an excited state in the interaction regime where the attractive polaron is the ground state. Remarkably, the molaron is stabilized  even in the regime of negative scattering lengths where in the model~\eqref{Eq:2-Ch_Hamiltonian} no molecule exists in vacuum, reminiscent of Cooper pairing in the theory of superconductivity~\cite{Cooper_1956,Bardeen_1957}. 

The single particle spectral function can be obtained via $\mathcal{A}(E,\pv) = \text{Im}\, G^R(E,\pv)$
and is shown for the molecule at unitarity, $a\rightarrow\infty$, in the left most plot of Fig.~\ref{fig:Fig1}(a), obtained using a non-selfconsistent $T$-matrix resummation approach. The molaron dispersion is visible in the spectrum. It shows a minimum at finite momentum~\cite{Trefzger_2012,Kamikado2017,Ness_2020,Schmidt_2011}, a finding robust with respect to the theoretical approximation scheme.

\textit{{Precursor of FFLO.---}} We now directly connect the existence of the molaron dispersion minimum at finite momentum  to the emergence of the elusive FFLO phase of superconductivity.
Extending the $T$-matrix approach to finite \textit{fermionic} impurity density, we keep track of the \textit{bosonic} molaron spectral function in dependence on the impurity chemical potential~$\mu_d$. As can be seen in Fig.~\ref{fig:Fig1}(a), the dispersion minimum continuously evolves until the finite-momentum molaron becomes  gapless precisely at the predicted onset of the FFLO phase at $\mu_d=0.196\,\epsilon_F$~\cite{Yoshida_2007}. This continuous relation of the molaron spectral function towards a gapless spectrum implies a simple picture of FFLO as a condensate of molarons.

To further substantiate this direct connection between molarons and the formation of the FFLO phase we investigate its dependence on interaction strengths. To this end we compare in Fig.~\ref{fig:Fig1}(b) the predicted quantum critical value of the FFLO transition for different mass ratios to the critical interaction strengths at which the minimum of the molaron dispersion moves to finite momentum.  As  can be seen, the boundaries lie in close proximity and exhibit the same behavior with respect to tuning of the range parameter $r^*$. Based on this close correspondence, the transition point towards FFLO can already be inferred from the excited composite states in the Fermi polaron problem. 

\textit{Molecular injection spectroscopy.---}
Despite its importance for emerging many-body phases, the molaron has not been observed in experiments at interaction strengths below the transition (see Fig.~\ref{fig:NewFig1}) \cite{Schirotzek_2009,Kohstall_2012,Sidler_2017,Ness_2020,Fritsche_2021}.
Here we propose a generalization of atomic Raman spectroscopy to allow for observing composite states including their full excitation spectrum. In Raman spectroscopy lasers induce  transfers between  internal atomic degrees of freedom adding a momentum $\vecq_L$ and an energy $\omega$ to the atoms \cite{Gotlibovych2014,Shkedrov2020,Ness_2020,Shkedrov2022}. 

Within linear response theory the absorption rate with respect to the perturbing atomic transition operator $\hat V_{\qv_L}$ is given by Fermi's Golden rule
\begin{align}
    \mathcal{R}(\omega,\vecq_L) = \sum_\alpha \left|\langle \alpha|\hat V_{\qv_L}|i\rangle \right|^2\delta(\omega-E_\alpha+E_i) 
    \label{Eq:FGR}
\end{align}
where  $\ket{i}$, $\{\ket{\alpha}\}$ and $E_i$, $E_\alpha$ denote the initial state and a basis set of final states and their respective energies.
 The polaron spectral function can be measured by choosing an initial state in which the impurity and the Fermi gas do not interact with each other. The transition operator then transfers the impurity to a final state in which it interacts with the fermions. In such injection spectroscopy the Raman response is identical to the polaron spectral function (see Ref.~\cite{Kohstall_2012} for rf-spectroscopy, $\qv_L=0$, which has been successfully employed to observe Fermi polarons). However, in this approach the molaron remains completely inaccessible due to a vanishing overlap between the initial
 state of fully delocalized particles and the final state where an impurity is fully localized around one of the bath fermions (cf. Fig. \ref{fig:NewFig1}). In ejection spectroscopy, on the other hand, an initially interacting impurity gets ejected into a non-interacting state. This technique allows only for the detection of the ground state. Moreover, finite-momentum properties of the molaron are not accessible.

We now show how the idea of injection spectroscopy can be extended to make the full excitation spectrum of molarons accessible. The idea can most easily be theoretically explained using a wave function picture.
To this end it is helpful to realize that the diagrammatic calculation leading to $G^R_\text{mol}$ in Fig.~\ref{fig:Fig1}(a) is equivalent to diagonalizing the problem in a truncated Hilbert space corresponding to a molaron wave function ansatz~\cite{Mora_2009,Punk_2009,Trefzger_2012}:
\begin{align}
 |M^{\vecp}\rangle=\alpha^{\pv}m_{\pv}^{\dagger}|\text{FS}_{N-1}\rangle+\sum_{\kv}\beta_{\kv}^{\pv}c_{-\kv}^{\dagger}d_{\kv+\pv}^{\dagger}|\text{FS}_{N-1}\rangle.
 \label{Eq:Ansatz_Molecule}
\end{align}
This ansatz is an extension of the vacuum solution and creates a molecule on top of a Fermi sea $|\text{FS}_{N-1}\rangle$ with $N-1$ atoms.

The vanishing overlap in injection spectroscopy highlights how the choice of an initial state is the key to measuring molecular properties. The state should fulfill two main criteria:
\begin{itemize}
    \item[(a)] It has to be a good reference state, i.e., a state that can be reliably prepared and whose  properties are  well-understood.
     \item[(b)] It should have sufficient spectroscopic overlap with the final state of interest, in this case the molecular state in the quantum medium.
\end{itemize}
We now show that starting from a relatively deeply-bound molecular state one can fulfill both criteria which allows to reliably probe many-body dressed composites in what we term \emph{molecular injection spectroscopy}. 
In this scheme, the first criterion is fulfilled by starting from a molecular state with binding energy $\epsilon_{B,\text{in}}/\epsilon_F\gg 1$ so that medium corrections determined by the Fermi energy $\epsilon_F$ are negligible. As a result, the initial state is well described by typical atomic physics models~\cite{chin:2010}. Establishing the fulfillment of the second condition  requires a  detailed analysis of the action of the Raman  operator $ \hat{V}_{\qv_L}$ on this initial state. 

Within our two-channel model, understanding the action of the Raman lasers requires translation of $\hat{V}_{\qv_L}$ from an atomic state basis (where it takes a form $\sim \sum_\pv  d_{\pv+\qv_L,f}^\dagger  d^{\phantom{\dagger}}_{\pv,i}$, with $i,f$ labeling the internal
atomic states of the impurity before and after the
Raman transition) into a basis that explicitly accounts for the closed-channel molecule $m^\dagger$ that arises from having integrated out atom fluctuations in the closed-channel. To achieve this we turn to an ab-initio coupled-channel calculation in the two-body limit. Considering here the two-body limit is justified since the initial state is tightly bound and many-body dressing of the final state molecule only affects its low energy physics, and hence does not affect the form of the laser operator.

\begin{figure}[t]
	\centering
	\includegraphics[width=
	\linewidth]{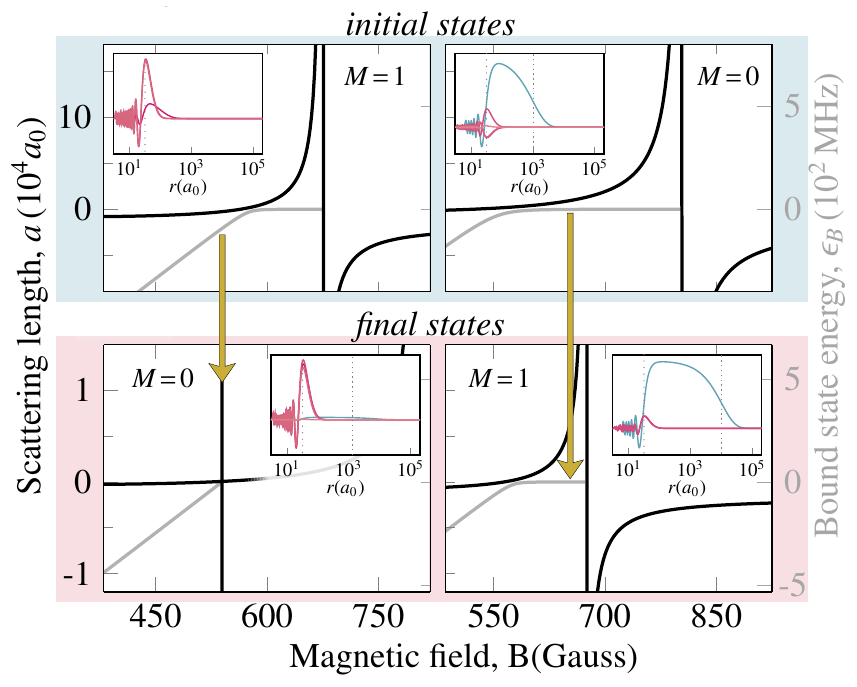}
	\caption{Experimental protocol for the example of $^6$Li. Scattering length (solid black) as a function of magnetic field for the initial (top, blue) and final state (bottom, red). In the left (right) panels one aims to probe a molaron in a final state at a narrow (broad) Feshbach resonance. The insets show the different channel contributions to the radial wave function of the Feshbach molecule at the applied magnetic field: Open channels (blue) and closed channels (pink). In the top right plot the narrow resonance is omitted for clarity. All data is obtained from a coupled-channel calculation, using realistic atomic potentials as input \cite{julienne:2014}. Dashed (dotted) lines in the insets indicate the van-der-Waals and scattering length, respectively. 
	}
	\label{fig:Fig2}	
\end{figure}

The ab-initio calculation is based on atomic states (see Appendix~\ref{App:CoulpedChannel}) and yields not only the binding energies and
the magnetic field dependent  scattering lengths, but also allows for a clear distinction between the open-channel (long-range component) and closed-channel (short-range component) contributions to the molecular wave functions~\footnote{Note that we refer here to an `open-channel' as the channel in which particles resides asymptotically in the scaterring process. Using this terminology a channel is also closed for a Feshbach molecule when it is below the collision threshold.}, see Fig~\ref{fig:Fig2}. For concreteness we consider here exclusively the example of $^6$Li, which features all key elements to demonstrate the idea of molecular injection spectroscopy. Specifically, we focus on
two limits where the initial state molecule
has either its weight almost entirely in the closed channels (cf. Fig. 3, upper left), or in
the open channel (upper right), allowing for a precise characterization of the Raman laser operator.

The left-hand panels in Fig.~\ref{fig:Fig2} show the scenario of a  strongly bound closed-channel Feshbach molecule in the initial state, corresponding to $|i\rangle=m^{\dagger}_{\mathbf{0},i}|\text{0}\rangle$ in our model. Such a state can be prepared for
sufficient detuning from a Feshbach resonance, possible for both narrow and broad resonances.
This choice of initial state (the wave functions including their hyperfine state contributions are shown as insets in Fig.~\ref{fig:Fig2}, see also Appendix~\ref{App:CoulpedChannel}) is ideal to detect molarons in the final state close to a narrow resonance, due to a large spectroscopic overlap between initial and final closed-channel contributions (compare insets in Fig.~\ref{fig:Fig2}). Furthermore, with a size on the order of the van-der-Waals length $l_{vdW}$ (dashed, vertical lines in insets), the initial closed-channel contributions yield only small overlap with the spatially extended open-channel states in the final state. As a consequence, the Raman operator is well approximated as $\hat V_{\qv_L} = \sum_\pv m_{\pv+\qv_L,f}^\dagger m^{\phantom{\dagger}}_{\pv,i\phantom{f}}\!\!$ in the two-channel model.

A second option, best suited to detect molarons close to a broad resonance in the final state, is to start from a deeply-bound initial molecular state close to an open-channel dominated resonance (see right hand panels in Fig.~\ref{fig:Fig2}). Note that, compared to the previous scenario, the initial state is less deeply-bound and in the regime where its energy does not depend linearly on the B-field.
As the coupled-channel calculation shows, the initial state is dominated by open-channel contributions. In the two-channel model this state is described by $\ket{i}=\sum_{\kv}\beta_\kv^{\mathbf{0}} c^\dagger_{-\kv} d^\dagger_{\kv,i}\ket{0}$. Thus the Raman laser mostly acts on that contribution which is confined on the order of the scattering length $a$ (dotted vertical line in the inset of Fig.~\ref{fig:Fig2}), and transferred to open-channel contributions in the final state manifold.
The projection onto closed-channels of the final state manifold
has a negligible contribution due to a lacking overlap of these states at low energy.  Hence, within the two-channel model the Raman operator is well represented by $\hat V_{\qv_L}=\sum_{\pv}d^{\dagger}_{\pv+\qv_L,f}d_{\pv ,i}^{\phantom{\dagger}}$.
 
\begin{figure}
	\centering
	\includegraphics[width=
	\linewidth]{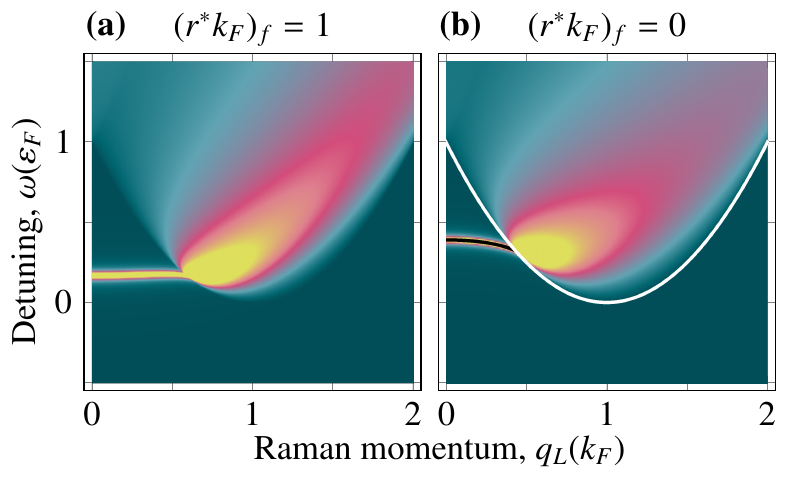}
	\caption{Raman spectra at unitarity and mass balance. \textbf{(a)} Narrow Feshbach resonance ($r^*k_F=1$) in the final state.
	The Raman spectrum equals the molaron spectral function. \textbf{(b)} Broad Feshbach resonance ($r^*k_F=0$) in the final state and initial state at $1/ak_F=1.5$. The Raman spectrum reproduces all key spectral features such as the onset of the continuum (white line) and sharp quasiparticle excitations (black line). Initial state energies are subtracted in both spectra.}
	\label{fig:Fig3}	
	
\end{figure}
\textit{Theoretical Raman spectra.---} 
Having established the form of the operator $\hat V_{\qv_L}$ we now turn to the prediction of the Raman absorption $\mathcal R(\omega,\qv_L)$. To this end, using the identity $\text{lim}_{y\rightarrow 0^+}1/(x+iy)=-i\pi \delta(x)+\mathcal{P}\left(1/x\right)$ and replacing $E_\alpha|\alpha\rangle=\mathcal{H}|\alpha\rangle$ we eliminate the explicit final state dependence in Eq.~\eqref{Eq:FGR}
\begin{align}
    \!\!\mathcal{R}(\omega,\vecq_L) \!=-\frac{1}{\pi}\, \text{Im}\!\left\langle i\left|\hat V_{\qv_L}\frac{1}{\omega-\mathcal{H}+E_i+i0^+}\hat V_{\qv_L}\right|i\right\rangle\!. \!\!
    \label{Eq:RamanSpectrum}
\end{align}
Here, the initial state $\ket{i}$ is given by a molecule state of form  \eqref{Eq:Ansatz_Molecule} with energy $\epsilon_{B,in}\gg \epsilon_F$ such that the many-body dressing by bath particles is negligible. 

Using a  basis truncation that includes up to one excitation on top of the Fermi sea, the Raman response takes the  form (for details see Appendix~\ref{App:ComputationRamanSpectra})
\begin{align}
     \mathcal{R}(\omega,\vecq_L)&=-\frac{1}{\pi}  \operatorname{Im}\Bigg( f^{\qv_L}(\omega)
G^{R}_{\text{mol}}(\qv_L,\omega)\Bigg)\nonumber \\
&+ \sum_{\kv} |\tilde{\beta}_{\mathbf{k}}^{\qv_L}|^2 \delta(\omega + E_{\text{mol},i} - E_{\text{FS}_{N-1}}\!\!- \varepsilon_\kv^c - \varepsilon_{\qv_L+\kv}^d),
\label{Eq:RamanSpecFullExpression}
\end{align}
where $\tilde{\beta}_\kv^{\qv_L}$ is the open-channel contribution of $\hat V_{\qv_L}\ket{i}$  and  $f^{\qv_L}$ is a multiplicative structure factor. 
Eq.~\eqref{Eq:RamanSpecFullExpression} shows the direct connection between the Raman response and the molaron Green's function $G^{R}_{\text{mol}}$. Exemplary spectra are shown in Fig.~\ref{fig:Fig3}.

In the case of a narrow Feshbach resonance in the final state (Fig.~\ref{fig:Fig3}(a)), we choose a deeply-bound initial state given by $\ket{i}=m^{\dagger}_{i,\mathbf{0}}\ket{\text{FS}_{N-1}}$. In this case, $\tilde{\beta}_\kv^{\qv_L}=0$ and $f^{\qv_L}\equiv1$,  such that the Raman spectrum and the molecular spectral function exactly coincide, $\mathcal{R}=\mathcal{A}_{\text{mol}}$.

Next we turn to a broad Feshbach resonance in the final state ($\ket{i}=\sum_{\kv}\beta_\kv^{\mathbf{0}} c^\dagger_{-\kv} d^\dagger_{\kv,i}\ket{\text{FS}_{N-1}}$). In Fig.~\ref{fig:Fig3}(b)  it can be seen that the Raman spectrum contains the same qualitative features as the corresponding molaron spectral function shown in Fig.~\ref{fig:Fig1}(a): in particular the position of the molaron as well as its merging into the continuum can be inferred from the Raman spectrum. Quantitatively, the difference between these spectra is merely a redistribution of spectral weight where the second part  of Eq.~\eqref{Eq:RamanSpecFullExpression} has negligible contribution~\footnote{It should be noted however that for certain observables such a redistribution is important such as when one aims to infer decay rates of quasiparticles from spectral functions.}. 
Importantly, the dispersion relation of the molaron including its finite momentum minimum is contained in such Raman spectra. Therefore, our approach allows one to observe a key signature of the instability towards FFLO both for broad and narrow Feshbach resonances. \\ 

\textit{Conclusion.}--- 
In this work we have presented a protocol to measure the momentum-resolved molaron spectral function  at arbitrary interaction strengths. This is achieved using Raman injection spectroscopy with a tunable transfer momentum, where the system is initialized in the ultrastrong coupling limit. The protocol allows for the first time the simultaneous observation of both polaron and molaron branches at the same interaction strength,
which provides an experimental tool to prove not only their coexistence but also the first order nature of their transition~\cite{Peng_2021,Parish_2021,Cui_2020}.
Our results show the robustness of this approach to observe the non-trivial dispersion relations of composite states, including the formation of a roton-type minimum. Furthermore, we demonstrated that this finite-momentum minimum in the  molaron spectral function is a precursor of the elusive FFLO phase. 

Our approach can be equally applied to the case of Bose polarons, where the resulting composite is fermionic. 
Such impurity systems hold promise to exhibit precursors of topologically non-trivial Fermi surfaces and  Fermi surface reconstruction.
Furthermore, it may allow to shed new light on the role of many-body bound states involving more than one bath atom as well as emerging phases in mass-imbalanced ultracold gases~\cite{Liu_2022}.

\textit{Acknowledgements.}---
We thank Yoav Sagi, Gal Ness and Selim Jochim for inspiring discussions. We acknowledge support by the Deutsche Forschungsgemeinschaft under Germany's Excellence Strategy -- EXC-2111 -- 390814868. This work has been supported by the Danish National Research Foundation through the Center of Excellence 'CCQ' (Grant No. DNRF156). O.~K.~D. and  J.~v.~M. are supported by  fellowships of the International Max Planck Research School for Quantum Science and Technology (IMPRS-QST).

\bibliography{biblio}
\appendix

 \section{Coupled-Channel calculation for two-body problem}
 \label{App:CoulpedChannel}
 In this appendix we provide supplementary information on the coupled-channel calculation leading to the results presented in the main text (see for example Ref.\ \cite{julienne:2014}). The aim is to find the scattering lengths, bound-state energies and wave functions for two interacting $^6\text{Li}$ atoms, labelled by indices $1$ and $2$, in an external magnetic field $B$. Neglecting magnetic dipole interactions and assuming zero rotational angular momentum ($s$-wave scattering), the Hamiltonian in the center of mass frame of this system is given by 
\begin{align}
    \hat{\mathcal{H}}&= \hat{\mathcal{H}}_{\text{Li}}(\hat{\mathbf{I}}_1,\hat{\mathbf{S}}_1, B) +\hat{\mathcal{H}}_{\text{Li}}(\hat{\mathbf{I}}_2,\hat{\mathbf{S}}_2, B) \nonumber\\
    &- \frac{1}{2\mu R}\frac{\partial^2}{\partial R^2}R + V(R, \hat{\mathbf{S}}_1,\hat{\mathbf{S}}_2).
\end{align}
Here  $\hat{\mathbf{I}}_i,\hat{\mathbf{S}}_i$ denote the nuclear and electronic spin operators of the two particles, $R$ is the interatomic distance, $\mu$ the reduced mass, $V$ the interaction potential and $\hat{\mathcal{H}}_{\text{Li}}$ the Hamiltonian of a free lithium atom in a magnetic field \cite{Gehm2003}. We use singlet and triplet interaction potentials \cite{julienne:2014} which have been optimized to match experiments. 

We can now express the wave function in terms of the asymptotic spin eigenbasis, denoted by $|i\rangle$, and a position basis in $R$. 
\begin{align}
    \ket{\Psi (R)}=\hat{P}_{\text{asym}} \frac{1}{R} \sum_i \psi_i(R)\ket{i}.
\end{align}
We have included a geometric factor $R$ in the definition of the radial wave function contribution $\psi_i$ to channel $i$ and  $\hat{P}_{\text{asym}}$ is the anti-symmetrization operator .  

In terms of the variables $\psi_i$, the problem now reduces to a second-order matrix-valued differential equation in $R$, which we solve with the renormalized Numerov method \cite{Johnson1978} with variable stepsize \cite{Vigo2005}. Since the total projection $M=\Io_1^z+\So_1^z+\Io_2^z+\So_2^z $ of the angular momentum on the magnetic field axis is conserved, the scattering/bound-state problem can be solved separately for every value of $M$. For every $M$ different combinations of the electronic and nuclear spins can contribute. An example of a channel in the $M=0$ manifold is $|m_{s_1}=1/2,m_{I_1}=1,m_{s_2}=-1/2,m_{I_2}=-1\rangle$. For ultracold collisions of ground-state atoms, one channel asymptotically lies below the scattering threshold (``open") and several lie above (``closed"). 

The radial wave functions are shown in Fig.~\ref{fig:Fig2}  for the $M=1$ and $M=0$ scattering manifolds of $^6\text{Li}$ for given magnetic field strengths. Here we have drawn the open channel in blue and the closed channels in pink. One can then compute the corresponding bound state energies, which are shown as grey lines in the main panels Fig.~\ref{fig:Fig2}. 
From the long-distance properties of scattering wave functions one can furthermore extract the corresponding scattering lengths, shown as black lines in Fig.~\ref{fig:Fig2}.

\section{Computation of Raman spectra}
\label{App:ComputationRamanSpectra}
In order to compute the Raman spectra defined in Eq.~\eqref{Eq:RamanSpectrum} of the main text, the matrix elements of the resolvant operator $(\rho-\mathcal{H})^{-1}$ need to be determined on the final state manifold spanned by states of the form $m_{\qv_L,f}^{\dagger}|\text{FS}_{N-1}\rangle$ and  $\{c_{-\kv}^{\dagger}d_{\kv+\qv_L,f}^{\dagger}|\text{FS}_{N-1}\rangle \}$ with 
$\rho= \omega+ E_i + i0^+$ and $|\kv|>k_F$. To this end, we rewrite the operator as
\begin{align}
 \frac{1}{\rho-\mathcal{H}}=\frac{1}{\rho-\epsilon}+ \frac{1}{\rho-\mathcal{H}}\,T\,\frac{1}{\rho-\epsilon}   
\end{align}
where $\epsilon$ and $T$ denote the kinetic and interaction terms of the Hamiltonian in Eq.~\eqref{Eq:2-Ch_Hamiltonian}, respectively. 
Defining 
\begin{align}
     \ket{0}&= m_{\qv_L,f}^{\dagger} \ket{\text{FS}_{N-1}}\\
    \ket{\kv}&= c_{-\kv}^{\dagger}d_{\kv+\qv_L,f}^{\dagger}\ket{\text{FS}_{N-1}}
\end{align} along with  
\begin{align}
\epsilon^0&=\xi_{\qv_L}+\nu+E_{\text{FS}}(N-1) \\
\epsilon_\kv &= \varepsilon_{\kv}^c+ \varepsilon_{\qv_L+\kv }^d+E_{\text{FS}}(N-1)
\end{align}
one arrives at the following system of equations:
\begin{align}
    \bra{0}\frac{1}{\rho-\mathcal{H}} \ket{0}&=  \frac{1}{\rho - \epsilon^0} +\frac{1}{\rho - \epsilon^0} \bra{0}\frac{1}{\rho -\mathcal{H} }\frac{h}{\sqrt{V}}\sum_{\kv} \ket{\kv} \nonumber\\
    \bra{0}\frac{1}{\rho-\mathcal{H}} \ket{\kv}&= \frac{1}{\rho - \epsilon_\kv} \bra{0}\frac{1}{\rho -\mathcal{H}} \frac{h}{\sqrt{V}} \ket{0}\nonumber\\
    \bra{\kv}\frac{1}{\rho-\mathcal{H}} \ket{0}&=  \frac{1}{\rho - \epsilon^0} \bra{\kv}\frac{1}{\rho -\mathcal{H} }\frac{h}{\sqrt{V}}\sum_{\kv'} \ket{\kv'}\nonumber \\
    \bra{\kv'}\frac{1}{\rho-\mathcal{H}} \ket{\kv}&=\frac{\delta_{\kv,\kv'}}{\rho -\epsilon_{\kv}}+  \frac{1}{\rho - \epsilon_\kv} \bra{\kv'}\frac{1}{\rho -\mathcal{H} } \frac{h}{\sqrt{V}} \ket{0} \ . 
\end{align}
This system can be solved and yields
\begin{align}
\bra{0}\frac{1}{\rho-\mathcal{H}} \ket{0}&=\frac{1}{h^2} \frac{1}{\frac{\rho - \epsilon^0}{h^2}-\frac{1}{V}\sum_{\kv''} \frac{1}{\rho - \epsilon_{\kv''}} }\nonumber\\
\bra{0}\frac{1}{\rho-\mathcal{H}} \ket{\kv}&=\frac{1}{\sqrt{V}h} \frac{1}{\rho - \epsilon_\kv}   \frac{1}{\frac{\rho - \epsilon^0}{h^2}-\frac{1}{V}\sum_{\kv''} \frac{1}{\rho - \epsilon_{\kv''}} }\nonumber\\
    \bra{\kv'}\frac{1}{\rho-\mathcal{H}} \ket{0}&= \frac{1}{\sqrt{V}h } \frac{1}{\rho -\epsilon_{\kv'}} \frac{1}{\frac{\rho - \epsilon^0}{h^2}- \frac{1}{V} \sum_{\kv''}\frac{1}{\rho -\epsilon_{\kv''}}  } \nonumber\\
            \bra{\kv'}\frac{1}{\rho-\mathcal{H}} \ket{\kv}&=\frac{\delta_{\kv,\kv'}}{\rho -\epsilon_{\kv}}\nonumber\\
        &+\frac{1}{V}  \frac{1}{\rho - \epsilon_\kv}\frac{1}{\rho -\epsilon_{\kv'}}   \frac{1}{\frac{\rho - \epsilon^0}{h^2}-  \frac{1}{V}\sum_{\kv''}\frac{1}{\rho - \epsilon_{\kv''}} }
\end{align}
where all sums are restricted to $|\kv''|> k_F$. 
As can be seen, the retarded molecular Green's function given by
\begin{align}
G^R(\omega,\qv_L)=\bra{0}(\rho-\mathcal{H})^{-1} \ket{0} 
\end{align}
reappears within  all other matrix elements of $(\rho-\mathcal{H})^{-1}$.
For two arbitrary overlapping Feshbach resonances, after acting on an initial state given by Eq.~\eqref{Eq:Ansatz_Molecule} in the main text with the Raman lasers the resulting state is given by
\begin{align}
 V_L^{\qv_L}|i\rangle=&\tilde{\beta}_0^{\qv_L}m_{\qv_L,f}^{\dagger}|\text{FS}_{N-1}\rangle+\sum_{\kv}\tilde{\beta}_{\kv}^{\qv_L}c_{-\kv}^{\dagger}d_{\kv+\qv_L,f}^{\dagger}|\text{FS}_{N-1}\rangle.
 \label{Eq:MoleculeAfterTransition}
\end{align}
Here the relative weights between the closed and open channel contribution $\tilde{\beta}_0^{\qv_L}, \tilde{\beta}_{\kv}^{\qv_L} $ can in general be different from the ones in Eq.~\eqref{Eq:Ansatz_Molecule}, as they depend on the form of the Raman laser operator. 

Finally, given knowledge of the $\tilde{\beta}_0^{\qv_L}, \tilde{\beta}_{\kv}^{\qv_L} $, the Raman response function of two arbitrary overlapping Feshbach resonances is given by 
\begin{align}
    \mathcal{R}^{\qv_L}(\omega)&=-\frac{1}{\pi}  \operatorname{Im}\Bigg(|\tilde{\beta}_{0}^{\qv_L}|^{2}\bra{0} \frac{1}{\rho-\mathcal{H}}\ket{0}\nonumber\\
    &+\sum_{\kv}2 \operatorname{Re}\left[\tilde{\beta}_{\mathbf{k}}^{\qv_L*} \tilde{\beta}_{0}^{\qv_L}\right]\bra{\kv} \frac{1}{\rho-\mathcal{H}} \ket{0}\nonumber\\
    &+\sum_{\kv \kv^{\prime}}\tilde{\beta}_{\kv'}^{\bar{\qv*}} \tilde{\beta}_{\mathbf{k}}^{\qv_L}\  \bra{\kv'}\frac{1}{\rho-\mathcal{H}}\ket{\kv}\Bigg).
\end{align}
Thus, it can easily be seen that with the exception of the (trivial) first term within $\bra{\kv'}\frac{1}{\rho-H} \ket{\kv}$ the resulting Raman response function contains the molecular Green's function, i.e. 
\begin{align}
    \mathcal{R}^{\qv_L}(\omega)=-\frac{1}{\pi}  \operatorname{Im}\Bigg( f^{\qv_L}(\omega)
G^{R}(\qv_L,\omega)\Bigg)\nonumber\\
+ \sum_{\kv} |\tilde{\beta}_{\mathbf{k}}^{\qv_L}|^2 \delta(\operatorname{Re}(\rho) -\epsilon_{\kv})
\label{Eq:RamanSpec}
\end{align}  
where the proportionality function is given by
\begin{align}
   f^{\qv_L}(\omega)= |\tilde{\beta}_{0}^{\qv_L}|^{2}
    +
    \sum_{\kv}2 \operatorname{Re}\left[\tilde{\beta}_{\mathbf{k}}^{\qv_L*} \tilde{\beta}_{0}^{\qv_L}\right]\frac{h}{\sqrt{V}}\frac{1}{\rho-\epsilon_\kv}\nonumber\\
    +
    \sum_{\kv \kv^{\prime}}\tilde{\beta}_{\kv'}^{\bar{\qv*}} \tilde{\beta}_{\mathbf{k}}^{\qv_L}\frac{h^2}{V}\frac{1}{\rho-\epsilon_\kv}\frac{1}{\rho-\epsilon_{\kv'}}.
\end{align}

In the main text, we discuss in detail the two scenarios of a narrow and a broad Feshbach resonance in the final state. In the first scenario, the laser operator takes the simple form $\hat V_{\qv_L} = \sum_\pv m_{\pv+\qv_L,f}^\dagger m^{\phantom{\dagger}}_{\pv,i}$ and due to the choice of initial state, this results in $\tilde{\beta}_0^{\qv_L}=\beta_0^{\qv_L}=1$ and $\tilde{\beta}_{\kv}^{\qv_L}=0$.  
Therefore, $f^{\qv_L}=1$ and the Raman spectrum~\eqref{Eq:RamanSpec} reduces exactly to the single particle spectral function. 

In the second scenario, the laser operator takes the form $\hat V_{\qv_L}=\sum_{\pv}d^{\dagger}_{\pv+\qv_L,f}d_{\pv ,i}^{\phantom{\dagger}}$. Therefore $\tilde{\beta}_{0}^{\qv_L}=0$ and due to our choice of initial state $\tilde{\beta}_{\kv}^{\qv_L}=\beta_{\kv}^{\qv_L}$, which can be simply obtained from minimizing the energy functional $\langle M^{\qv_L}|\mathcal{H}-E|M^{\qv_L}\rangle$ and is given by
\begin{align}
|\beta_{\kv}^{\qv_L}|^2=\frac{1}{V}&\left(\frac{1}{\omega+i0^+ -\varepsilon_{\kv}^c-\varepsilon_{\kv+\qv_L}^d-E_{\text{FS}}(N-1)}\right)^2\nonumber\\
&\frac{1}{\frac{1}{h^2}+\frac{1}{V}\sum_{\kv'}\big(\frac{1}{\omega+i0^+-\varepsilon_{\kv'}^c-\varepsilon_{\kv'+\qv_L}^d-E_{\text{FS}}(N-1)}\big)^2}.
\end{align}

\end{document}